\documentclass[aps,pre,twocolumn,floatfix,superscriptaddress,showpacs,showkeys]{revtex4-1}
\usepackage{epsf,amsmath,amssymb,verbatim,color,multirow,pifont,mathrsfs,soul,amsthm, wasysym}
\usepackage{graphicx, upgreek}
\usepackage[us,12hr]{datetime}

\begin{document}
\today

\title{Percolation critical exponents in cluster kinetics of pulse-coupled oscillators}
\author{Gangyong Gwon}
\affiliation{Department of Physics, Jeonbuk National University, Jeonju 54896, Republic of Korea}

\author{Young Sul Cho}\email{yscho@jbnu.ac.kr}
\affiliation{Department of Physics, Jeonbuk National University, Jeonju 54896, Republic of Korea}
\affiliation{Research Institute of Physics and Chemistry, Jeonbuk National University, Jeonju 54896, Republic of Korea}

\begin{abstract}
Transient dynamics leading to the synchrony of pulse-coupled oscillators has previously been studied as an aggregation process of synchronous clusters, and a rate equation for the cluster size distribution has been proposed.
However, the evolution of the cluster size distribution for general cluster sizes has not been solved yet.
In this paper, we study the evolution of the cluster size distribution from the perspective of a percolation model
by regarding the number of aggregations as the number of attached bonds.
Specifically, we derive the scaling form of the cluster size distribution
with specific values of the critical exponents using the property that the characteristic cluster size diverges as the percolation threshold is approached from below. Through simulation, it is confirmed that the scaling form well explains the evolution of the cluster size distribution. 
Based on the distribution behavior, we find that a giant cluster of all oscillators is formed discontinuously at the threshold and also that further aggregation does not occur like in a one-dimensional bond percolation model. Finally, we discuss the origin of the discontinuous formation of the giant cluster from the perspective of global suppression in explosive percolation models. For this, we approximate the aggregation process as a cluster--cluster aggregation with a given collision kernel.
We believe that the theoretical approach presented in this paper can be used to understand the transient dynamics of a broad range of synchronizations.
\end{abstract}

\maketitle

\section{Introduction}
The synchronization of oscillators is a phenomenon in which oscillators evolve to the same state by interaction among themselves~\cite{sync_review}.
Such phenomena are observed in diverse real systems and have been widely studied~\cite{sync_review2, sync_review3, sync_neural, sync_neural2, motter_powergrid, kuramoto_powergrid}. 
In particular, pulse-coupled oscillators refer to oscillators linked by sudden pulses rather than continuous interaction~\cite{peskin_pulse, mirollo_pulse, pulse_coupled1, pulse_coupled2}.
Pulse-coupled oscillators are observed in real systems such as flashing fireflies~\cite{firefly1, firefly2}, firing neurons~\cite{fire_neuron1, fire_neuron2}, and others~\cite{pulse_coupled_other1, pulse_coupled_other2, pulse_coupled_other3}.
To understand the behaviors of these systems, various models have been proposed and studied.

One such model of pulse-coupled oscillators, so-called scrambler oscillators, was studied in~\cite{sync_agg}. This model is the stochastic version of traditional models
that obey deterministic resetting rules~\cite{peskin_pulse, mirollo_pulse}.
In this model, a fixed number $N$ of oscillators are given, and each oscillator indexed by $i$ for $i=1,...,N$ has a voltage $x_i(t) \in [0, 1]$ at time $t \geq 0$, where each set of oscillators having the same $x_i$ is called a (synchronous) cluster. Initially at $t=0$, each $x_i$ is given randomly in the range of $[0, 1]$, and thus all oscillators belong to isolated clusters. Following the dynamical rules below, all oscillators become synchronized and form a single cluster.
\begin{itemize}
\item[(i)] Each $x_i$ increases linearly according to $\dot{x}_i=1$.
\item[(ii)] When $x_i$ of a cluster of $j$ oscillators reaches the threshold value of $1$, the cluster fires and scrambles every other cluster to have a new random voltage in the range of $[0, 1]$.
\item[(iii)] The firing cluster of size $j$ absorbs any other clusters with voltages in the range of $[1-j/N, 1)$ 
by bringing them to the threshold and synchronizing with them.
\item[(iv)] The voltage of the firing cluster resets to $0$ along with those of the absorbed clusters.
\end{itemize}
We note that all oscillators in the same cluster have the same value of the new voltage in (ii). Therefore, all oscillators in the same cluster sustain synchronization, meaning that no cluster is broken after it is formed. As a result, this model can be interpreted as an irreversible cluster aggregation, where 
clusters group to form larger ones following (iii).

In~\cite{sync_agg}, cluster densities $c_s(t)$ denoting the number of clusters of size $s$ over $N$ at time $t$ were studied.
With the assumptions that the cluster voltages are uniformly distributed and that the cluster densities are of different sizes and uncorrelated,   
the rate equation for $c_s(t)$ was obtained. Using the rate equation, closed forms of $c_1(t)$, $c(t)=\sum_sc_s(t)$, and
moments $M_k(t) = \sum_s s^k c_s(t)$ for $k=2,3,4$ were derived.  
Then, $c_s(t)$ for small values of $s \geq 2$ were obtained through numerical integration of the rate equation. However, obtaining $c_s(t)$ for $s \gg 1$ in this manner is not feasible because all combinations of cluster sizes with sums equal to $s$ should be considered to calculate the gain term of $\dot{c}_s$ following rule (iii) above.
Consequently, the evolution of $c_s$ for general values of $s \gg 1$ has remained unexplored.

In this paper, we analyze the aggregation of synchronous clusters of scrambler oscillators from the perspective of a percolation model. 
This approach allows us to apply the scaling ansatz for $c_s$, which has been well established in the study of percolation models.
As a result, we establish the scaling form of $c_s$ that well describes the evolution of $c_s$ for general values of $s \gg 1$. 
We then obtain various percolation properties using the scaling form of $c_s$, thereby providing a general understanding of the system.

The rest of this paper is organized as follows. In Sec.~\ref{sec:perspective_PT},
we introduce a new parameter corresponding to the bond density of a percolation model and discuss the relation between the parameter and time $t$. In Sec.~\ref{sec:critical_exponent}, we analyze the cluster aggregation of scrambler oscillators as a phenomenon in a percolation model using the new parameter. Then we obtain the percolation critical exponents observed in the cluster aggregation. In Sec.~\ref{sec:Kij_i},
the underlying mechanism for the discontinuous formation of a giant cluster of scrambler oscillators is revealed by approximating the aggregation process as a cluster--cluster aggregation with a given collision kernel. In Sec.~\ref{sec:conclusion}, we summarize the results and discuss related future works.

\section{Interpretation via cluster aggregation in a percolation model}
\label{sec:perspective_PT}

In a percolation model, two clusters are aggregated into one when a bond is attached between them, and the evolution of $c_s$ is studied as a function of bond density. Following this convention, we study the evolution of $c_s$ of scrambler oscillators as a function of $p$ corresponding to bond density, where $p$ is defined as the number of aggregations over $N$. Following this definition, $p=0$ is given initially and $p$ increases by $1/N$ whenever a cluster is absorbed to the firing cluster following rule (iii)~\cite{scramble_detail}, which continues up to $p=1-1/N$. Moreover, $c$ decreases by $1/N$ as $p$ increases by $1/N$ because the total number of clusters decreases by one as two clusters aggregate. Since the total number of clusters is $N$ and thus $c=1$ at $p=0$, we obtain $c(p) = 1-p$ for $0 \leq p \leq 1-1/N$. We note that
the upper bound of $p$ is $1$ in the thermodynamic limit $N \rightarrow \infty$.

\begin{figure}[t!]
\includegraphics[width=0.8\linewidth]{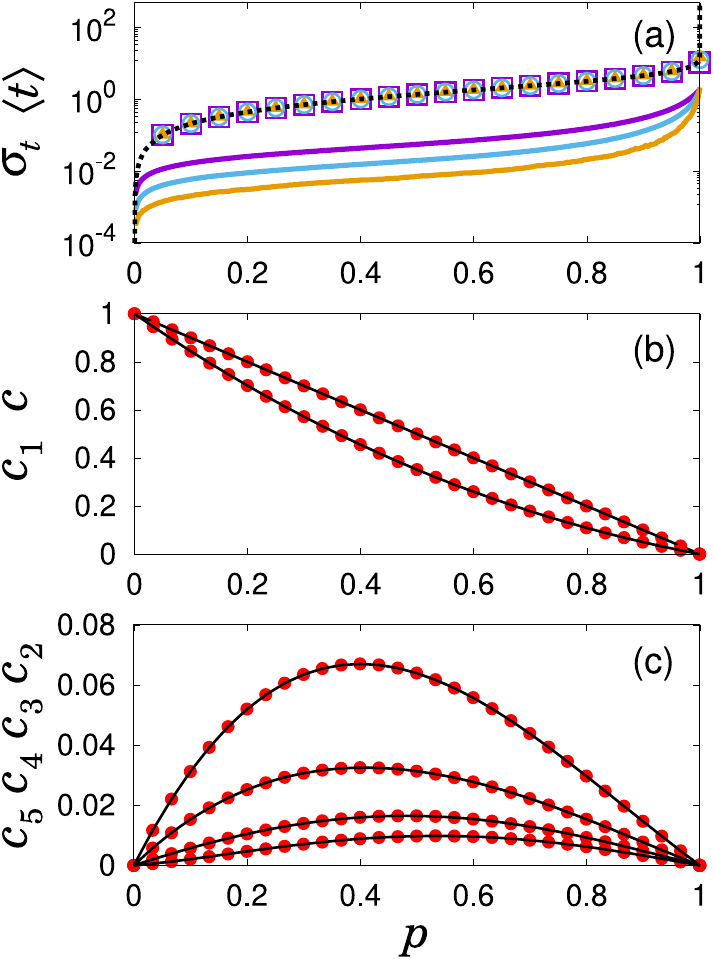}
\caption{(a) Ensemble average $\langle t \rangle$ and the standard deviation $\sigma_t$ of $t$ for each value of $p$. 
Symbols denote $\langle t \rangle$ for $N/10^3 = 2^1\,(\square), 2^{4}\,(\bigcirc), 2^{7}\,(\blacktriangle)$,
and the dotted line is Eq.~(\ref{eq:t_vs_p}).
Solid lines denote $\sigma_t$ for $N/10^3=2^1,\,2^4,\,2^7$ from top to bottom.
(b) Plot of $c$ and $c_1$ obtained using simulation $(\bullet)$ from top to bottom. Solid lines are theoretical curves $c=1-p$ and $c_1(p)$ obtained by solving Eq.~(\ref{eq:rate_scramble}) with $s=1$. (c) Plot of
$c_s$ for $s=2,3,4,5$ obtained using simulation $(\bullet)$ from top to bottom. Solid lines are theoretical curves obtained by solving Eq.~(\ref{eq:rate_scramble}) numerically with $s=2,3,4,5$ from top to bottom.} 
\label{Fig:t_ci_vs_p}
\end{figure}

We compare $c(p)=1-p$ with $c(t)=\text{exp}(-t/2)$ introduced in~\cite{sync_agg} such that we derive the relation between $p$ and $t$ as
\begin{equation}
p = 1 - \text{exp}\Big(-\frac{t}{2}\Big).
\label{eq:t_vs_p}
\end{equation}
We verify this relation by showing that the ensemble average of $t$ for a fixed value of $p$ follows Eq.~(\ref{eq:t_vs_p}) and that its standard deviation decreases to $0$ as $N \rightarrow \infty$ through simulation, as shown in Fig.~\ref{Fig:t_ci_vs_p}(a).
We note that $t$ monotonically increases with $p$, where $t(p=0)=0$ and $t \rightarrow \infty$ as $p \rightarrow 1$.

From now on, we study the behaviors of $c_s(p)$ and other quantities such as $M_k(p) = \sum_s s^k c_s(p)$ and the threshold $p_c$ for the emergence of a giant cluster 
according to the parameter $p$ using the theoretical framework of a percolation model.
We remark that these results correspond to the results at the original time $t=-2\text{ln}(1-p)$ by the relation in Eq.~(\ref{eq:t_vs_p}),
which means that our result provides a general understanding of the original system.

\section{Derivation of critical exponents}
\label{sec:critical_exponent}
\subsection{Critical exponent of the $k$-th moment of the cluster size distribution}
\label{sec:gamma_k}

In this section, we obtain the critical exponent $\gamma_k$ for $M_k(p) \propto (p_c-p)^{-\gamma_k}$ 
with $p_c=1$ for general values of $k \geq 2$.
We obtain the rate equation for $c_s(p)$ by inserting $t = -2\textrm{ln}(1-p)$ 
into the rate equation for $c_s(t)$ introduced in~\cite{sync_agg} as follows,
\begin{equation}
\frac{dc_s}{dp} = -\frac{2c_s}{c} + \sum_{i=1}^{s}\frac{c_i}{c}e^{-ic}\sum_{\sum_j ja_j = s-i} \Bigg(\prod_{j \geq 1}\frac{(ic_j)^{a_j}}{a_j!}\Bigg),
\label{eq:rate_scramble}
\end{equation}
where the rightmost summation is over
all combinations of $\{a_j\}$ for non-negative integers $a_j \geq 0$ satisfying $\sum_j ja_j=s-i$.
We confirm that $c_s(p)$ obtained through the integration of Eq.~(\ref{eq:rate_scramble}) and 
obtained through simulation agree well for $1 \leq s \leq 5$, as shown in Fig.~\ref{Fig:t_ci_vs_p}(b) and 1(c).

We then introduce the generating function $G(z,p)=\sum_{s \geq 1}c_s(p)e^{sz}$, which is used to derive $M_k(p)$ using the relation $M_k(p)=\partial^kG(z,p)/\partial z^k|_{z = 0}$. We multiply $e^{sz}$ and sum over $s \geq 1$ on both sides of Eq.~(\ref{eq:rate_scramble}) such that we obtain
\begin{equation}
(1-p)\frac{\partial G(z, p)}{\partial p} + 2G(z,p) = G(x, p),
\label{eq:rate_scramble_generating}
\end{equation}
where $x \equiv z-(1-p)+G(z,p)$ and $x \rightarrow 0$ as $z \rightarrow 0$ by the relation $G(z=0, p)=1-p.$ 
We obtain $dc/dp = -1$ from Eq.~(\ref{eq:rate_scramble_generating}) using $G(z=0,p) =c$ with $c = 1-p$, as expected.

We apply $\partial^k/\partial z^k$ to both sides of
Eq.~(\ref{eq:rate_scramble_generating}) and substitute $z=0$ and $x(z=0)=0$.
We take the asymptotic behavior of $M_k(p) \propto (1-p)^{-\gamma_k}$ as $p \rightarrow 1$, because $p_c = 1$ as discussed later (see Sec.~\ref{sec:tau_sigma}) and $M_k(p)$ diverges as $p \rightarrow p_c$. 
Then we consider the dominant terms on both sides as
$p \rightarrow 1$ and obtain
\begin{equation}
(1-p)\frac{\partial M_k(p)}{\partial p} = (2^k-1)M_k(p)
\label{eq:Mk_dominant}
\end{equation}
for $k \geq 2$.
Finally, we obtain $\gamma_k = 2^k-1$ for $k \geq 2$ by inserting
$M_k(p) \propto (1-p)^{-\gamma_k}$ to both sides of
Eq.~(\ref{eq:Mk_dominant}).
[For a more detailed derivation of Eq.~(\ref{eq:Mk_dominant}) from Eq.~(\ref{eq:rate_scramble_generating}), see Appendix A.]
We check that $\gamma_k=2^k-1$ is valid for $k=2,3,4$ via simulation in Fig.~\ref{Fig:scramble_M} and via comparison with closed forms of $M_k(p)$ in Appendix A.

\begin{figure}[t!]
\includegraphics[width=0.8\linewidth]{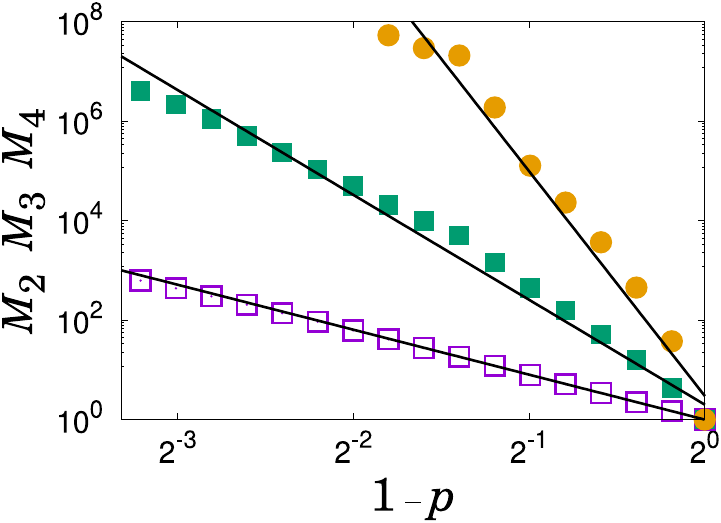}
\caption{Results of $M_k(p)$ obtained by performing simulation for $k=2$ $(\square)$, $k=3$ $(\blacksquare)$, and $k=4$ $(\bullet)$ with $N = 128,000$. Solid lines are $(1-p)^{-\gamma_k}$ with $\gamma_k = 2^k-1$.} 
\label{Fig:scramble_M} 
\end{figure}

\subsection{Critical exponents $\tau$ and $\sigma$ of the cluster size distribution}
\label{sec:tau_sigma}

\begin{figure}[t!]
\includegraphics[width=0.8\linewidth]{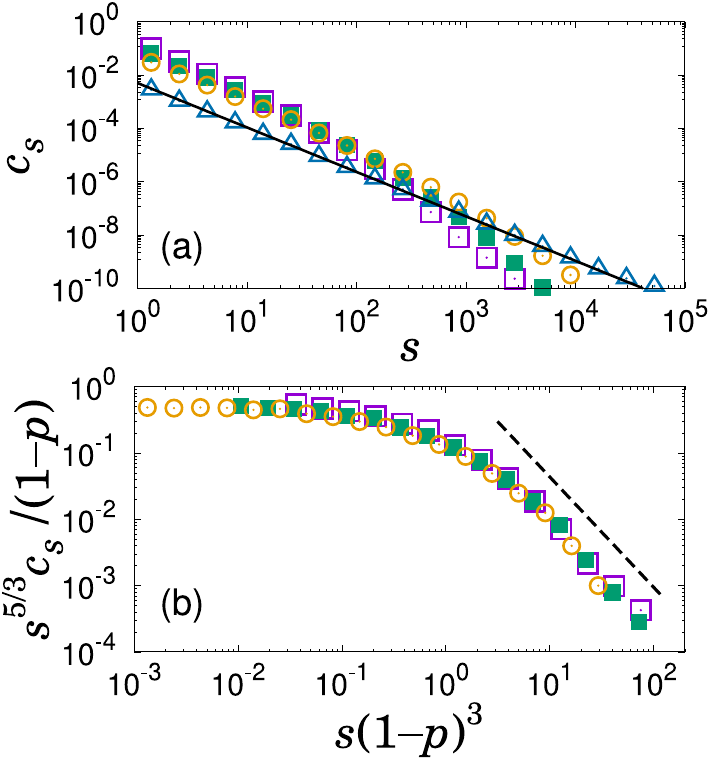}
\caption{(a) Results of $c_s(p)$ obtained by performing simulation for $p=0.7\,(\square), 0.8\,(\blacksquare), 0.9\,(\bigcirc), 0.99\,(\triangle)$ with $N = 128,000$. The slope
of the solid line is $-5/3$. (b) Data collapse of $s^{5/3}c_s(p)/(1-p)$ vs.
$s(1-p)^3$ for $c_s(p)$ with the $p=0.7,\,0.8,\,0.9$ presented in (a). The slope of the dashed line
is $-5/3$.} 
\label{Fig:scramble_cdist}
\end{figure}

In this section, we derive the critical exponents of $c_s$.
In Fig.~\ref{Fig:scramble_cdist}(a), we observe $c_s \propto s^{-\tau}$ with $1 < \tau < 2$ at $p$ values close to $1$.
By taking the conventional scaling theory of $c_s$ as $p$ approaches $p_c$ from below, we assume that $c_s = As^{-\tau}$ for $1 \leq s \leq s_{\xi}$ with an $s$-independent factor $A$ and also that $c_s$ decays rapidly for $ s > s_{\xi}$, for which the characteristic cluster size $s_{\xi} \propto (p_c-p)^{-1/\sigma}$
and $1<\tau<2$.

We consider the subcritical region $p < p_c$ where no giant cluster exists and thus $\sum_{s=1}^{\infty}sc_s = 1$.
Then we obtain $A \propto (p_c-p)^{(2-\tau)/\sigma}$ using $1 = \sum_{s=1}^{\infty}sc_s \approx A\int_1^{s_{\xi}}s^{1-\tau}ds$.
We note that $A$ depends on $p$ unlike the case when $A$ is constant for $\tau > 2$ as observed in second-order percolation transitions~\cite{stauffer, can}.
This allows us to obtain $p_c=1$ and $\tau+\sigma=2$ from the relation $(1-p)=c(p)=A(p)\sum_{s=1}^{\infty}s^{-\tau} \propto (p_c-p)^{(2-\tau)/\sigma}$
as $p \rightarrow p_c$ from below, for which we use the property that $\sum_{s=1}^{\infty} s^{-\tau} = \zeta(\tau)$ is finite for $1 < \tau < 2$.
We note here that $\tau$ derived in this way satisfies the condition $1 < \tau < 2$ self-consistently, as shown in the next paragraph.

Based on the preceding discussions, we suggest the scaling form of $c_s$ of scrambler oscillators as
\begin{equation}
c_s = (1-p)s^{-\tau}f[s(1-p)^{1/\sigma}]
\label{eq:cs_scalingform}
\end{equation}
with $\tau + \sigma = 2$. 
Using this scaling form, we obtain the scaling relation $\gamma_2 = (3-\tau)/\sigma - 1$ from
$M_2 \approx (1-p)^{1-(3-\tau)/\sigma}\int_0^{\infty}\bar{s}^{2-\tau}f(\bar{s})d\bar{s}$ with $\bar{s} \equiv s(1-p)^{1/\sigma}$.
Here, we assume that $\int_0^{\infty}\bar{s}^{2-\tau}f(\bar{s})d\bar{s}$
is finite.   
In Sec.~\ref{sec:gamma_k}, we obtained $\gamma_2 = 3$ by solving Eq.~(\ref{eq:Mk_dominant}) for $k=2$.
By combining the two equations $3 = (3-\tau)/\sigma-1$ and $\sigma+\tau=2$, we obtain $\sigma=1/3$ and $\tau=5/3$. In Fig.~\ref{Fig:scramble_cdist}, we confirm that the scaling form in Eq.~(\ref{eq:cs_scalingform}) with $\sigma = 1/3$ and $\tau = 5/3$
is valid using simulation data.

We now extend the above to obtain the relation $\gamma_k = (1+k-\tau)/\sigma - 1$ for arbitrary $k \geq 3$ using the scaling form Eq.~({\ref{eq:cs_scalingform}}) under the assumption that $\int_0^{\infty}\bar{s}^{k-\tau}f(\bar{s})d\bar{s}$ is finite. However, the scaling relation 
$\gamma_k = (1+k-\tau)/\sigma - 1$ contradicts the exact result $\gamma_k = 2^k-1$ obtained by solving Eq.~(\ref{eq:Mk_dominant}) in Sec.~\ref{sec:gamma_k}. 
To resolve this contradiction, we surmise that $f(\bar{s})$ would not decrease exponentially but rather decrease in a polynomial manner as $\bar{s}$ increases, such that $\int_0^{\infty}\bar{s}^{k-\tau}f(\bar{s})d\bar{s}$ diverges for $k \geq 3$
and $\gamma_k = (1+k-\tau)/\sigma - 1$ does not hold for $k \geq 3$.
To support this, we check that $f(\bar{s}) \propto \bar{s}^{-\tau}$ with $\tau = 5/3$ for $\bar{s} \gg 1$ in Fig.~\ref{Fig:scramble_cdist}. We find that $\int_0^{\infty} \bar{s}^{k-\tau}
f(\bar{s})d\bar{s}$ diverges for $k \geq 3$ while it is finite
for $k=2$, as expected.

\begin{figure}[t!]
\includegraphics[width=1.0\linewidth]{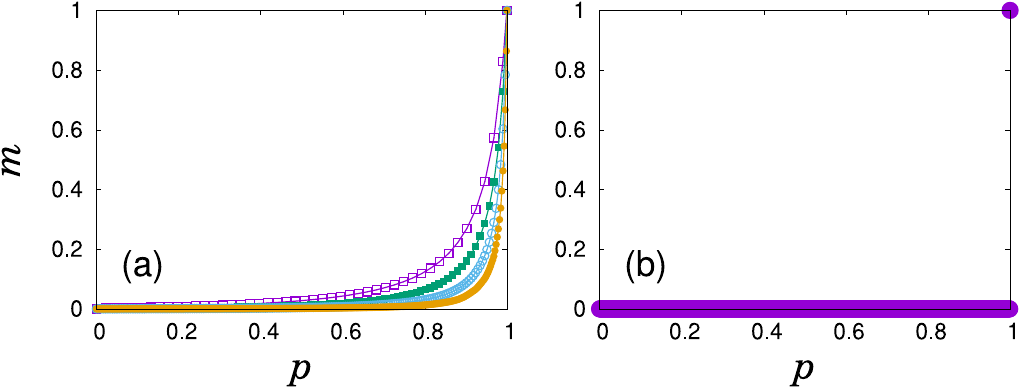}
\caption{(a) Giant cluster size $m$ as a function of $p$ for 
$N/10^3 = 2\,(\square), 8\,(\blacksquare), 32\,(\circ), 128\,(\bullet)$ from the left,
where $m$ increases more drastically near $p=1$ as $N$ increases.
(b) Schematic diagram for $m$ vs. $p$ in the thermodynamic limit $N \rightarrow \infty$.} 
\label{Fig:m_vs_p}
\end{figure}

\section{Underlying mechanism for the discontinuous formation of a giant cluster}
\label{sec:Kij_i}

In this section, we discuss the cluster aggregation of scrambler oscillators from the perspective of a percolation transition. Since $0 \leq p \leq 1$, $p_c = 1$ means that no giant cluster
exists for $p < 1$ but a giant cluster containing all oscillators emerges at $p=1$. This is consistent with the idea that every cluster contains a small number of oscillators up to the very large $t$ region, which was mentioned in~\cite{sync_agg}. If $m$ denotes the fraction of oscillators belonging to the giant cluster, $m$ behaves as 
\begin{flalign}
m =
\begin{cases}
0 & \text{if}~~p<1, \\\\
1 & \text{if}~~p=1.
\end{cases}
\label{eq:m_scrambler}
\end{flalign}
We verify this behavior of $m$ through simulation, as shown in Fig.~\ref{Fig:m_vs_p}.
As a result, a giant cluster of scrambler oscillators is formed discontinuously as in one-dimensional bond percolation~\cite{stauffer}
and real systems such as diffusion-limited cluster aggregation~\cite{disc_real}.

In previous studies on explosive percolation models~\cite{ep, ep_dorogovtsev}, it was shown that a giant cluster is formed discontinuously when the growth of large clusters is globally suppressed~\cite{riordan, jan, cho_science, ep_growing, hybrid}. 
To demonstrate that the same mechanism is also involved in the aggregation process of scrambler oscillators, we propose a simplified model that describes the aggregation process in an approximate manner, and then at the end of this section we discuss how the above mechanism fits into the proposed model.

With scrambler oscillators, $n \geq 0$ number of clusters can be absorbed by a firing cluster for a given $c_s(p)$, after which $p$ is increased by $p \rightarrow p + n/N$. If the size of the firing cluster is $s$, then the probability that the number of absorbed clusters is $n$ is given by $(cs)^n e^{-cs}/n!$. Therefore, the average number of clusters absorbed to the firing cluster of size $s$ is $\sum_{n \geq 0} n(cs)^n e^{-cs}/n!= cs$. We remark that the average number of clusters absorbed by the firing cluster is proportional to the size of the firing cluster. 
Finally, the average number of aggregations for each firing is $\langle n\rangle=\sum_{s \geq 1} (c_s/c)cs = 1$, because the probability that the size of a firing cluster is $s$ is $c_s/c$.

\begin{figure}[t!]
\includegraphics[width=0.8\linewidth]{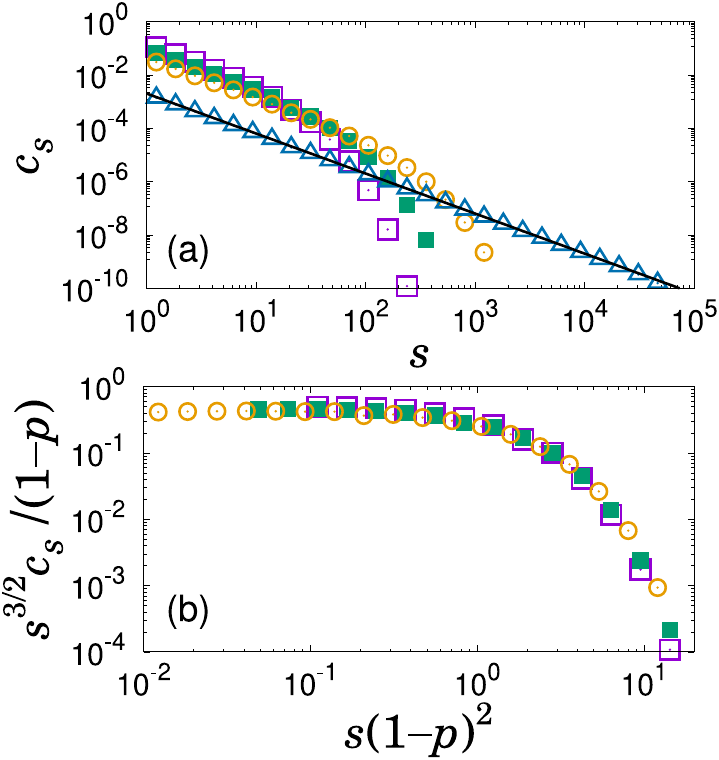}
\caption{Results of cluster--cluster aggregation with the collision kernel $K_{ij} = i$. (a) Symbols are $c_s(p)$ obtained by performing simulation for $p=0.7\,(\square), 0.8\,(\blacksquare), 0.9\,(\bigcirc), 0.995\,(\triangle)$ with $N=128,000$. The slope
of the solid line is $-3/2$. (b) Data collapse of $s^{3/2}c_s(p)/(1-p)$ vs.
$s(1-p)^2$ for $c_s(p)$ with the $p=0.7,\,0.8,\,0.9$ presented in (a).} 
\label{Fig:smol_Kij_i_cdist}
\end{figure}

Based on these properties, we approximate the aggregation process of scrambler oscillators with a cluster--cluster aggregation, where two clusters of sizes $i$ and $j$ are chosen with a probability proportional to $K_{ij} = i$ for each aggregation $p \rightarrow p+1/N$. 
For the cluster--cluster aggregation with the collision kernel $K_{ij} = i$, we derive the scaling form of $c_s$ written in Eq.~(\ref{eq:cs_scalingform}) with $\tau=3/2$ and $\sigma=1/2$. We note that $\tau=3/2$ is similar to $\tau = 5/3 \approx 1.66$ of the scrambler oscillators, which reflects that both processes have similar behaviors of aggregating cluster sizes.
We check that $c_s(p)$ for different values of $p$ collapse onto a single curve using the scaling form, as shown in Fig.~\ref{Fig:smol_Kij_i_cdist}.
Derivation of these results is presented in Appendix B.

We now briefly discuss how the growth of large clusters is globally suppressed to cause the discontinuous formation of a giant cluster when $K_{ij}=i$.
For comparison, we consider the Erd\"{o}s--R\'enyi (ER) model~\cite{er}
in which a giant cluster is formed continuously at $p_c<1$ with $\tau > 2$.
The ER model reflects cluster--cluster aggregation with the collision kernel $K_{ij}=ij$, and thus 
two clusters of sizes $i$ and $j$ are chosen with the probability proportional to $ij$ for each aggregation $p \rightarrow p+1/N$.
Therefore, the growth of large clusters for $K_{ij}=i$ is globally suppressed compared to that in the ER model,
where the suppression is global because $K_{ij}=i$ is applied to all clusters. 
This suppression makes the cluster sizes similar, resulting in $1 < \tau < 2$ at $p_c$. 
Following the discussion for $1 <\tau < 2$ in Sec.~\ref{sec:tau_sigma}, it is naturally derived that a giant cluster is formed discontinuously 
at $p_c = 1$~\cite{can}.

\section{Conclusion}
\label{sec:conclusion}

In summary, we considered the evolution of the cluster size distribution $c_s$ of scrambler oscillators according to the number of aggregations following the convention of a percolation model in which the evolution of $c_s$ is studied according to the number of attached bonds. 
As a result, we established the scaling form Eq.~(\ref{eq:cs_scalingform}) with $\sigma=1/3$ and $\tau=5/3$ 
that describes the $c_s$ evolution for general values of $s \gg 1$.
When $1<\tau < 2$, the giant cluster size jumps to $1$ at $p_c=1$ discontinuously like that in one-dimensional bond percolation.
We then approximated the aggregation process to a cluster--cluster aggregation with the collision kernel $K_{ij} = i$
and discussed that global suppression of the growth of large clusters reveals a discontinuous formation of a giant cluster, like in explosive percolation models.

In this study, we considered the scrambler oscillators introduced in~\cite{sync_agg} to investigate the aggregation of clusters in the transient dynamics leading up to synchrony.
Future works can apply the analysis presented in this paper to studies of other pulse-coupled oscillators such as variants of scrambler oscillators having nonlinear charging curves~\cite{sync_agg2}.
We note that the theoretical framework for the evolution of the cluster size distribution proposed in this paper was established based on the divergence of the characteristic cluster size 
at the threshold, which is a general property of percolation models.
Accordingly, the present results may be used to understand a broad range of transient dynamics of synchronous clusters that can be described as irreversible cluster aggregations, 
such as aggregation in cluster synchronization as revealed by network symmetry~\cite{symmetry, pecora_sciadv_2016, yscho_prl_2017, remote_prl_2013, pecora_ncomm_2014} 
and aggregation in a path to global synchronization by increasing coupling strength~\cite{path_to_sync, structure_syook}.

\section*{Appendix A: Derivation of Eq.~(\ref{eq:Mk_dominant}) from Eq.~(\ref{eq:rate_scramble_generating})}
\label{sec:derivation_scrambler_gamma}

We apply $\partial^k/\partial z^k$ to both sides of Eq.~(\ref{eq:rate_scramble_generating}). It is straightforward to obtain $(1-p) \partial[\partial^k G(z,p)/\partial z^k]/\partial p + 2\partial^k G(z,p)/\partial z^k$ on the left side; however, 
derivation of the right side is not straightforward because $G(x,p)$ has $x\equiv z-(1-p)+G(z,p)$ in place of $z$. We obtain the general form of $\partial^k G(x, p)/\partial z^k$ as
\begin{equation}
\frac{\partial^k G(x,p)}{\partial z^k} = \sum_{k'=1}^{k}\frac{\partial^{k'}G(x,p)}{\partial x^{k'}}g_{k, k'}(z,p),
\label{eq:Mk_derivation_G}
\end{equation}
where $g_{k, k'}$ can be derived using the recurrence relation
\begin{flalign}
g_{k+1, k'}=
\begin{cases}
\frac{\partial x}{\partial z}g_{k, k} & \text{if}~~k'=k+1, \\\\
\frac{\partial x}{\partial z}g_{k, k'-1}+\frac{\partial g_{k, k'}}{\partial z} & \text{if}~~2 \leq k' \leq k, \\\\ 
\frac{\partial g_{k, 1}}{\partial z} & \text{if}~~k'=1,
\end{cases}
\label{eq:Mk_derivation_g}
\end{flalign}
for $k \geq 1$ beginning with $g_{1,1}=\partial x/\partial z$. 
Here, $g_{k,k}=(\partial x/\partial z)^k$ and $g_{k, 1} = \partial^k x/\partial z^k$
can be derived using Eq.~(\ref{eq:Mk_derivation_g}). Therefore, by applying $\partial^k/\partial z^k$ to both sides of Eq.~(\ref{eq:rate_scramble_generating}), we obtain
\begin{eqnarray}
&&(1-p)\frac{\partial}{\partial p}\Big[\frac{\partial^k G(z,p)}{\partial z^k}\Big] + 2\frac{\partial^k G(z,p)}{\partial z^k} \notag\\ 
&=& \frac{\partial G(x,p)}{\partial x}\frac{\partial^k x}{\partial z^k} + \frac{\partial^k G(x,p)}{\partial x^k}\Big(\frac{\partial x}{\partial z}\Big)^k \notag \\
&+&\sum_{k^{'}=2}^{k-1}\frac{\partial^{k^{'}} G(x,p)}{\partial x^{k^{'}}}g_{k, k^{'}}(z,p)
\label{eq:rate_scramble_generating_partial}
\end{eqnarray}
for $k \geq 2$.

We substitute $z=0$ and $x(z=0)=0$ to Eq.~(\ref{eq:rate_scramble_generating_partial})
such that we obtain
\begin{equation}
(1-p)\frac{\partial M_k}{\partial p} = (2^k-1)M_k +\sum_{k^{'}=2}^{k-1} M_{k^{'}}g_{k,k^{'}}(z=0,p)
\label{eq:rate_scramble_Mk_partial}
\end{equation}
for $k \geq 2$, where we used $\partial G/\partial x|_{x=0}=1$, $\partial x/\partial z |_{z=0} = 2$, and $\partial^k x/\partial z^k |_{z=0}=M_k$ for $k \geq 2$. If we assume that $(2^k-1)M_k$ is the dominant term on the right side as $p \rightarrow 1$, by taking the asymptotic behavior of $M_k(p) \propto (1-p)^{-\gamma_k}$ 
as $p \rightarrow 1$, we obtain $\gamma_k = 2^k-1$ for $k \geq 2$ as we have already derived using Eq.~(\ref{eq:Mk_dominant}).

We now briefly demonstrate that $(2^k-1)M_k$ is indeed the dominant term on the right side as $p \rightarrow 1$
if we insert $M_k(p) \propto (1-p)^{-\gamma_k}$ with $\gamma_k = 2^k-1$ into Eq.~(\ref{eq:rate_scramble_Mk_partial}). 
In this case, each term $M_{k^{'}}g_{k,k^{'}}(z=0, p)$ for $2 \leq k^{'} \leq k-1$ in the summation on the right side diverges as 
$M_{k^{'}}g_{k,k^{'}}(z=0, p) \propto (1-p)^{-(\gamma_{k^{'}}+\gamma_{k-k^{'}+1})}$, where we used the recurrence relation
$g_{k+1, k^{'}}(z=0, p) \propto  (1-p)^{-\gamma_{k-k^{'}+2}+\gamma_{k-k^{'}+1}} g_{k, k^{'}}(z=0, p)$ obtained from
$g_{k+1, k^{'}}(z=0, p) \propto \partial g_{k, k^{'}}/\partial z|_{z=0}$ in Eq.~(\ref{eq:Mk_derivation_g})
for $k \geq k^{'}$ with $\gamma_1 = 0$ and $g_{k^{'}, k^{'}} = (\partial x/\partial z)^{k^{'}}$.
Dominant exponents of terms of the summation given by $\gamma_{k^{'}}+\gamma_{k-k^{'}+1} = 2^{k^{'}}+2^{k-k^{'}+1}-2$ for $2 \leq k^{'} \leq k-1$
are always smaller than
$\gamma_k = 2^k-1$. As a result, the first term $(2^k-1)M_k$ on the right side is dominant, which supports that $\gamma_k = 2^k-1$ for $k \geq 2$.

We obtain closed forms of $M_k(p)$ as  
\begin{flalign}
M_{k}(p)=
\begin{cases}
(1-p)^{-3} & \text{if}~~k=2, \\\\
7(1-p)^{-7} - 6(1-p)^{-6} & \text{if}~~k=3, \\\\ 
\frac{143}{10}(1-p)^{-15} - \frac{224}{5}(1-p)^{-10} \\ + \frac{63}{2}(1-p)^{-9} & \text{if}~~k=4,
\end{cases}
\label{eq:Mk_derivation_ex}
\end{flalign}
where $\gamma_k = (2^k-1)$ is checked for $k=2,3,4$.

\section*{Appendix B: Derivation of critical exponents for $K_{ij}=i$}
\label{sec:derivation_Kij_i}

The rate equation for $c_s(p)$ of the cluster--cluster aggregation with $K_{ij} = i$ is given by
\begin{equation}
\frac{d c_s}{dp} = -sc_s - \frac{c_s}{c} + \sum_{i=1}^{s-1}ic_i\frac{c_{s-i}}{c}.
\label{eq:rate_Kij_i}
\end{equation}
We multiply $e^{sz}$ and sum over $s \geq 1$ on both sides of Eq.~(\ref{eq:rate_Kij_i}) and 
use the generating function $G(z,p)=\sum_{s\geq1}c_s(p)e^{sz}$ such that we obtain
\begin{equation}
\frac{\partial G(z, p)}{\partial p} = -\frac{\partial G(z, p)}{\partial z} - \frac{1}{c}G(z, p) + \frac{1}{c}\frac{\partial G(z,p)}{\partial z}G(z, p),
\label{eq:rate_Kij_i_generating}
\end{equation}
where $dc/dp = -1$ is checked using $G(z=0,p) = c$ and $\partial G(z,p)/\partial z|_{z=0}=1$.

At first, we derive the critical exponent $\gamma_k$ for $M_k(p) \propto (1-p)^{-\gamma_k}$. 
To obtain $M_k(p)$ using the relation $M_k(p) = \partial_z^k G(z,p)|_{z = 0}$,
we apply $\partial_z^k$ to both sides of Eq.~(\ref{eq:rate_Kij_i_generating}) and substitute $z=0$.
For $k=1$, $d M_1/d p = 0$ is obtained using $G(z=0, p)=c$ and $M_1 = 1$, which is self-consistent.

For $k \geq 2$, we obtain
\begin{equation}
\frac{\partial M_k}{\partial p} = \frac{1}{c} \sum_{k'=1}^{k-1}\binom{k}{k'}M_{k'+1}M_{k-k'},
\label{eq:rate_Kij_i_moments}
\end{equation}
where we used $\partial_z^k [(\partial_z G) G] = \sum_{k^{'}=1}^{k-1}\binom{k}{k'}\partial_z^{k^{'}+1}G\partial_z^{k-k^{'}}G$.
We substitute $M_k(p) \propto (1-p)^{-\gamma_k}$ for some constants $\gamma_k > 0$ depending on $k$ into Eq.~(\ref{eq:rate_Kij_i_moments}).
Then we assume the relation $\gamma_k = \gamma_{k'+1} + \gamma_{k-k'}$ for $1 \leq k' \leq k-1$
by comparing the exponents of $(1-p)$ on both sides
and obtain the recurrence relation $\gamma_k = \gamma_{k-1} + \gamma_2$ for $k \geq 2$ using $k^{'} = 1$. By solving Eq.~(\ref{eq:rate_Kij_i_moments}) for $k=2$, we obtain
$M_2(p)=(1-p)^{-2}$ and thus $\gamma_2 = 2$. Using the recurrence relation with $\gamma_2=2$, we obtain $\gamma_k = 2(k-1)$, where the assumption $\gamma_k = \gamma_{k'+1} + \gamma_{k-k'}$ for $1 \leq k' \leq k-1$ is satisfied self-consistently.

Next, we derive the critical exponents $\tau$ and $\sigma$ of the scaling form
of $c_s$ given by Eq.~(\ref{eq:cs_scalingform}). By inserting Eq.~(\ref{eq:cs_scalingform}) into the relation $M_k = \int_{1}^{\infty} s^k c_s ds$ and comparing the exponents of both sides,
we obtain the relation $\gamma_k = k/\sigma + (1-\tau)/\sigma - 1$. Comparing this relation with $\gamma_k = 2(k-1)$ derived in the previous paragraph,
we obtain $\tau = 3/2$ and $\sigma = 1/2$.

\section{ACKNOWLEDGMENTS}
This work was supported by a National Research Foundation of Korea (NRF) grant, No. 2020R1F1A1061326.


\begin{thebibliography}{99}
\bibitem{scramble_detail} More than one clusters may be absorbed to the firing cluster in (iii). In this case, we treated those to be absorbed one by one in random order.
\end{thebibliography}
\end{document}